# *Softwarized 5G Networks Resiliency with Self-Healing*

José Sánchez[12], Imen Grida Ben Yahia[1], Noël Crespi[2], Tinku Rasheed[3], Domenico Siracusa[3]
[1] Orange Labs, Paris, France
[2] Telecom SudParis, Evry, France
[3] CREATE-NET, Trento, Italy

*Abstract*— The meaning of 5G is still a subject of discussion in the industry. However, the softwarization of networks is expected to shape the design, operation and management of 5G networks. The opportunity is then crucial for Telcos, vendors and IT players to consider the management of 5G networks during its design time and avoid the "build it first, manage it later" paradigm. However, network softwarization comes with its own set of challenges, including robustness, scalability and resilience. In this paper, we analyze the vulnerabilities of SDN (Software-Defined Networks) and NFV (Network Function Virtualization) from a fault management perspective, while taking into account the autonomic principles. In particular, we focus on resiliency and we propose a Self-Healing based framework for 5G networks to ensure services and resources availability.

## I. Introduction

Massive and large markets of applications are expected by 2020, such as the Internet of Things, Machine to Machine, public safety applications, and all the future yet unknown services that will need a very flexible network infrastructure. On the other side, new trends are characterizing the network transformation: the future "IT-ization" of the networks.

In the context of 5G, a customer expects services everywhere, every time and through any access technology. To fulfill this strong expectation, it implies to hide the notion of distance, latency, availability, etc., which assumes powerful mechanisms to master dynamic resources provisioning according to customers' needs, among other criteria [1].

What "*5G*" certainly means is still a subject of debate within the Telco industry. However, the softwarization or the IT-ization of networks is expected to be among the major factors to shape the 5G network design: In fact, the softwarization is an opportunity to rethink the network architecture as well as the corresponding management. More particularly, a key aspect to ensure the "best experience", "service availability" and the "reliability" is the resiliency of 5G networks. In this regard, we discuss a Self-Healing framework addressing faults within the so-called Softwarized networks; Software Defined Networking (SDN) and Network Function Virtualization (NFV) in the context of 5G.

In this paper, we give an overview of Self-Healing systems in section II and we analyze the impact of softwarization trends on future 5G networks in section III. In section IV, we propose a fault management framework, based on Self-Healing, for 5G SDN-based networks while looking beyond SDN and considering NFV architectures and analyzing how this Self-Healing may manage this architecture. Section V concludes the paper and highlights the future work.

## II. Self-Healing in a nutshell

Self-healing is an autonomic property [1] that recovers a managed system from abnormal states and maintains it in a healthy state. Self-Healing properties are survivability, availability, reliability, maintainability, stabilization, fault-tolerance, and human-assistance if needed. The Self-Healing architecture takes the shape of a control-loop mechanism composed of three blocks, namely, detection, diagnosis, and recovery, which receives information from the managed elements through sensors and acts upon them through the actuators.

A self-healing system analyzes two kinds of data: alarms that contain the state of the network elements to detect failures (broken states) or performance metrics of resources to measure degradations (degraded states). These metrics are various: availability, performances or resiliency of services, network functions and resources. Metrics can be general (delay, jitter or throughput), technology-specific (e.g. the rate of reply of the controller to the switches' requests in SDN). In case of failure or degradation, the diagnosis block determines the root cause and launches the proper recovery algorithm to address that specific problem. This execution is translated into reconfiguration or re-adaptation actions on the managed system to change its behavior.

To fulfill the strong requirements of 5G networks, we need an intelligent and autonomic fault management framework. We consider that a pro-active Self-Healing framework can forecast future service failures by considering the degradations on services' parameters and network elements and can propose recovery actions before the service failure occurs, preventing any service outage rather than acting once the failure has already occurred, like in traditional approaches that are prone to errors.

### III. IMPACT OF SOFTWARISATION IN 5G CONTEXT

The networking industry and the IT players are converging. This convergence is foreseen since long time ago and is now possible through the advances in the IT and networking industries: SDN and the NFV. Although SDN and NFV are independent, their global benefits are the leverage of service delivery, the reduction of costs, the enablement of 3rd party partnerships through standardized APIs, and the increase of network programmability versus the classical configurable networks.

SDN is a design approach, with several definitions: the separation of the data and the control plane, and the programmability of network elements in all networking domains through advanced network abstraction and independently from any southbound protocol. ONF (Open Networking Foundation) defines [2] SDN as a layered architecture composed of the data plane, the control plane, the application plane, as well as a transversal management plane whose implementation, inner blocks and functionalities are still fuzzy.

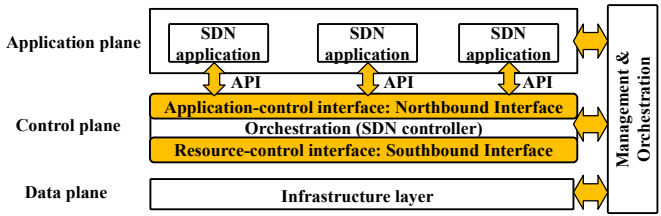

Fig. 1. SDN layered architecture

The application plane is composed of SDN applications that act as clients of the network services of the control plane. The control plane exposes Application Programming Interfaces (API) in order for the SDN applications to program the control plane through the northbound interface (e.g. RESTCONF API, REST API).

The application plane programs the data plane through the SDN controller. The control plane communicates with the data plane through the southbound interface protocol—whose most extended protocol is OpenFlow—. The separation between data and control plane consists in adopting a higher-level namespace and exploiting a logically-centralized controller to enforce the network policies by communicating them to generic forwarding hardware (via the southbound interface) in terms of language rather than technology-specific encoding.

NFV is a networking initiative led by Telcos [3], to replace Network Functions (e.g. Firewall, load balancing, cipher, etc.) —currently implemented in sophisticated and expensive hardware that hinders the integration with Telcos equipment— by software embedded in commodity hardware (high-volume standard servers, storage and switches). This approach reduces power consumption, what is very in line with 5G, allows for sharing of resources, reducing maintenance costs and time to deploy new functions/services. With this approach, Network Functions become Virtual Network Functions (VNF). ETSI NFV ISG defines a reference framework for NFV [3] (Fig. 2).

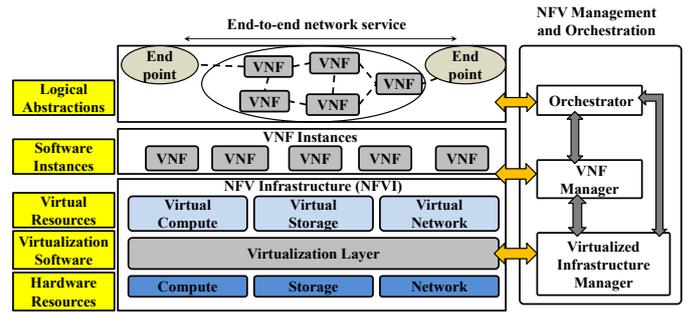

Fig. 2. NFV infrastructure [3]

The Management and Orchestration block is transversal to the OSS/BSS (Operations Support Systems/ Business Support Systems) and is composed of the orchestrator, the VNF Manager, and the Virtualized Infrastructure Manager. The orchestrator receives a data model description of the service from the OSS/BSS, depicted as a VNF Forwarding Graph, to be instantiated and it calculates the available resources to map this descriptor into physical locations. The VNF Manager ensures the migration of the current deployed VNFs to avoid any unavailability at all times. This block is in charge of migrating the VNFs.

Some work exists concerning attempts to virtualize network functions in both core networks—the work in [4] analyzes the functions of S-GW and P-GW EPC nodes and classifies them according to data plane and control plane, and access network—whereas the authors in [5] proposes virtualizing LTE network functions. Also, this work [6] presents a virtualized routing function over an OpenFlow network.

5G networks would benefit from SDN and NFV. Thus, it will inherit the vulnerabilities and issues known so far on both NFV and SDN paradigms.

In this regard, we study and identify hereafter the vulnerabilities in SDN-based 5G networks:

***Scalability:*** As the number of switches connected to the SDN controller augments, the requests of the network elements to the controller can congest the control links. The protocols acting on the west-east interfaces (WE) to manage the interconnection among different controllers' domains could improve the scalability, but it lacks investigation so far.

***Security:*** The brain of the SDN network—the controller— may be attacked and the whole network at its behest can be compromised. Indeed, SDN architecture is vulnerable to DoS attacks (Denial-Of-Service) that send many packets to the SDN controller to saturate it and leave the network inoperative.

***Resiliency:*** Scalability and security impact on the resiliency of the controller, undoubtedly the weakest point of this architecture. We will focus on this third point.

Despite these vulnerabilities of SDN, several gaps in the state of the art seem to be unnoticed. We propose a fault management framework that sheds light on these issues.

A few works explicitly address the management challenges of SDN, but most of them focus on the OpenFlow protocol, leaving out legacy equipment. They mainly propose traffic engineering recovery solutions that reconfigure the switches to forward through alternative paths to avoid the affected switches, but they omit the diagnosis part to a great extent. Malfunctions in the control plane are not adequately addressed [7], with the exception of [8], which considers failures on the controller and proposes a protocol to give the control to a backup controller.

We identify also a lack of proposals concerning integrated fault management frameworks for both OpenFlow-based and non-OpenFlow or legacy equipment—such as programmable eNodeBs (Evolved Node for LTE/UMTS), legacy switches and routers, no matter which southbound protocol they use (e.g. SNMP, NETCONF, etc.). Nevertheless, some interesting work combines SDN-like control for OpenFlow equipment or Ethernet (e.g. Dynamic flow routing) with traditional management for legacy equipment [9].

Most resiliency mechanisms for SDN focus on one of the two types of control in SDN networks, in-band or out-of-band. Even, little information is provided about the functioning of these types of control. Thus, depending on the type of control, the resiliency aspects and robustness differ to a great extent. In out-of-band control, the control traffic is transported apart in a separated control network—the controller communicates with the switches through dedicated control links. In the in-band control, data traffic and control traffic are intertwined in the same interfaces. Out-of-band solution is better in terms of isolation—a failure on a data link on an in-band controlled network affects control and data traffic, whilst a failure on a data link on out-of-band controlled networks does not affect the control traffic, as each network element has a separated control link to receive the rules from the controller—and less congested—both types of traffic (control and data) are intertwined within the in-band control—. Most resiliency mechanisms so far are proposed for in-band controlled networks [10]. With regard to this, we propose an Autonomic Fault Management mechanism, based on Self-Healing, that can manage both types of controlled networks.

Considering the virtualization of network functions in 5G through NFV, the following issues need also to be solved: given the high traffic demand in 5G and the number of devices to provision and to configure, one issue is to identify the VNF to be activated, the physical hosts and the VNF allocation to provide a given service. The key benefits from VNF are: the need to consider them dynamically provisioned and on demand allocation/configuration.

The combination of SDN and NFV for 5G networks results in generic hardware boxes and software to be managed across network segments. Importantly, what operators seek is the decoupling of hardware and software lifecycle management in order to reduce costs. The target is to program software to be tuned or adapted. In 5G, besides the physical complex infrastructure, a set of VMs will run on generic hardware through hypervisors. This will enable advanced configuration and customization of the network functions.

The aforementioned vulnerabilities of SDN and NFV will impact 5G management and highlight the need to rethink the current FCAPS (Fault, Configuration, Accounting, Performance, and Security) to ensure the reliability, availability and faster service delivery.

IV. SELF-HEALING FRAMEWORK FOR 5G NETWORKS

Autonomic management is agreed as the next generation of management. The first standard of global autonomic architecture is already defined by ETSI/NTECH [11]. Today, it is also part of the mobile network through the Self Organized Networks (SON) standardized by 3GPP [12] and is expected to evolve and reach further industrial adoption through the industrial driven research and innovation within the 5G Private Public Partnership [13].

Due to the strong requirements and complexity of 5G networks, we need to adopt NFV and SDN principles to benefit from the advantages of softwarisation. However, we are obliged to overcome their inherent vulnerabilities. For that, we follow autonomic principles like the Self-X functions, particularly Self-Healing systems, to address them. We undoubtedly need a Self-Healing system to monitor proactively the appropriate metrics of different layers and network segments to control those vulnerabilities and prevent any impact on services. Our contribution is then to ensure the resiliency and availability of end-to-end services in NFV-based architectures that rely on centralized SDN out-of-band and in-band controlled networks.

We propose a Self-Healing framework (Fig. 3) to ensure the availability of end-to-end services. It acts in the three planes of SDN and in the service plane, by taking observations from the network and launching recovery actions. It interacts with the SDN architecture by: (1) reconfiguring the end-to-end services, (2) orchestrating the SDN applications dynamically to face malfunctions or changing conditions, (3) reprogramming the data plane through the SDN controller, and (4) directly acting on the data plane to set specific configurations on legacy equipment or carrying out manual installation that the controller cannot perform by itself.

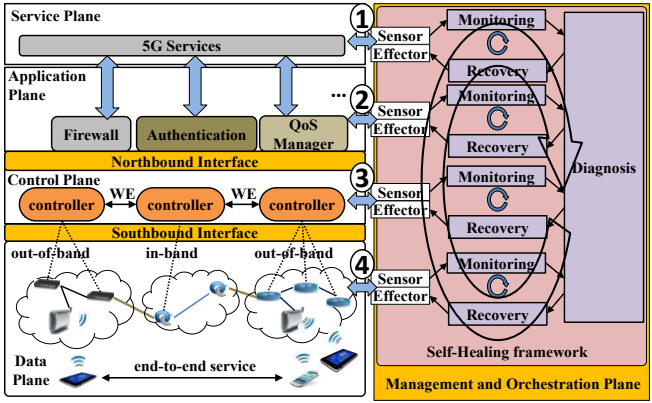

Fig. 3. Self-Healing framework for SDN

End-to-end network services of the service plane rely on SDN applications that are chained dynamically thanks to SDN to provide the service. This Self-Healing framework is composed of a Self-Healing control-loop per plane—with their corresponding three blocks—. Each control-loop retrieves symptoms from each plane and correlates them with the symptoms coming from the rest of planes. We provide with some examples of monitor actions per plane (Table I). For instance, under a failure on one service, we retrieve symptoms from the application plane, the control plane, the data plane and we correlate them. The application plane's symptoms and the data plane's discard any failures on the SDN applications and at a physical level, but the control plane's symptoms confirm that the root cause is due to the removal of rules installed on certain switches that are involved in the end-to-end service.

TABLE I. EXAMPLE OF CONTROL LOOPS ACTIONS IN EACH PLANE

| Plane | Task |
|---|---|
| Service | Lists the running services<br>Monitors the status of each network service<br>Lists the users using the service<br>Lists the Forwarding Graph of SDN applications for each service |
| Application | Monitors the state of applications<br>Monitors the allocated path of each application |
| Control | Monitors the controller(s)<br>Monitors the control links<br>Lists the managed switches by each controller across different domains |
| Data | Monitors the status of switches<br>Monitors the data links<br>Lists the flows/rules in each switch and their operation mode (standalone or secure)<br>Monitors the status of clients<br>Monitors the status of servers and hosts |

As SDN ensures the connectivity among the VNFs, acting as enabler for NFV, our Self-Healing framework can be used to manage the NFV architecture and its related vulnerabilities. NFV boosts greatly the programmability and flexibility by introducing a degree of granularity that SDN cannot provide. It introduces the orchestration of VNFs, by flexibly increasing the capacity of instantiated VNFs according to load (scale-up), instantiating as many VNFs as needed (scale-out), or migrating the current VNFs to avoid service interruptions. SDN, on the other hand, flexibly programs the network to connect together the underlying VNFs of the end-to-end service in a dynamical manner.

We envisage two types of recovery actions for our Self-Healing framework in NFV-SDN based networks: 1) recovery actions that heal the SDN architecture and 2) actions that cooperate with the NFV infrastructure to avoid any service interruption by upgrading, scaling or migrating the underlying VNFs. Table II depicts examples of both types of recovery actions on SDN-NFV networks. A Self-Healing system can intermediate between the NFV orchestrator and the VNF Manager to propose dynamic migrations, creations or suppressions of VNFs in response to failures on the SDN network. It could act as a central brain that orchestrates actions on SDN-NFV networks according to the diagnosed problem.

TABLE II. POSSIBLE RECOVERY ACTIONS APPLIED TO THE DIFFERENT PLANES

| Plane | Root cause | Recovery action |
|---|---|---|
| Service | End-to-end service misconfiguration | 1) Reconfigure end-to-end service<br>2) Reconfigure involved SDN applications |
| Service | Crash of end-to-end service | 1) Reinitiate involved SDN applications<br>2) Restart end-to-end service |
| Application | Crash of SDN application | 1) Restart application<br>2) Migration to other VM |
| Application | Too many requests to SDN application | 1) Instantiate other VM to carry out this application<br>2) Augment memory and CPU of VM |
| Application | Misconfiguration of SDN application | Configure SDN application |
| Control | Control link failure (in-band control) | 1) Set standalone mode on affected switches<br>2) Reconfigure switches to avoid the affected switch |
| Control | Control link failure (out-of-band control) | Set standalone mode on affected switches |
| Control | Controller failure | Balancing to a secondary controller |
| Data | Bridge misconfigured | Bridge reconfiguration |
| Data | High interference level | Reduction of uplink power on Access Point |
| Data | Misconfiguration on client application | Application reconfiguration |
| Data | Switch ignores how to reach client | Installation of flow on switch |

As example, in an end-to-end service composed of four chained VNFs (Fig. 4) whose locations may change dynamically across the network, we define the service topology as the subset of the network topology that contains the VNFs used by that service. In the presence of malfunctions that affect the VNFs, they are migrated to other physical locations to avoid the service interruption, what changes the service topology. The SDN controller dynamically reallocates the path to chain the four VNFs.

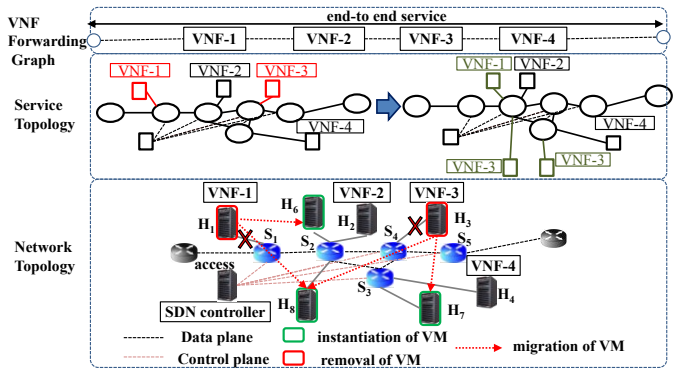

Fig. 4. Service topology changes in NFV-SDN architectures

By adopting SDN-NFV principles on 5G networks, we add more complexity on the diagnosis process as the service topology becomes dynamic due to the VNF migrations on the core network. We need a Self-Healing framework to be aware of both topological changes: 1) on the service topology of the core network, and 2) on the radio access network due to the clients' connections and disconnections on the move. The Self-Healing framework is aware of the latest healthy network state, which is updated every time the service topology changes (due to any of both reasons) and it estimates topology changes at run-time by comparing the latest healthy state to the current network state, provided by the SDN controller, and feeds the diagnosis engine with this information.

We translated part of this idea of Self-Healing framework into a specific SDN platform [15] by embedding a diagnosis block, based on Bayesian Networks, in the control plane. The SDN platform was composed of a POX controller connected to an in-band controlled and fixed network implemented on Mininet. We tested the diagnosis block in a multicast video service. We proved that our diagnosis block can detect any unavailability of the multicast video service or its underlying resources and can reactively resolve malfunctions at several levels: physical failures on the network elements, crashes on the application plane or malfunctions on the control plane.

## V. CONCLUSIONS AND FUTURE WORK

Today's networks are characterized by an overwhelming growth in data traffic due to the high adoption rate of smart phones and devices (with strong capacity of computing, storage and networking), cloud platforms and the new business models and "the always connected" users.

In this paper, we describe how 5G will emphasize those aspects with the "7 trillion devices for 7 billion people" vision and the expected shift from the "IP-ization to the IT-ization".

We expect 5G to be a more complex environment, imposing criteria and requirements, complexity and volume of data which can hardly be handled by traditional management schemes. Cognitive and autonomic management is a powerful vision; a promising solution that paves the way towards fully controllable and manageable sets of systems. We zoom on a possible Self–Healing framework for a SDN-based 5G network.

We highlight that the "IT-zation" is an opportunity to move from "configurable" to "programmable" networks, where the abstraction of networking protocols is mandatory to avoid the silo management operations and the basic CLI commands. Also, this transition encourages the rethinking the 5G network architecture itself.

In this work, we have focused on metrics that measure the availability of the services, network functions, and resources (physical and logical) involved in the delivery of those services over 5G SDN-based networks. The aim is to extend this Self-Healing reactive framework to consider other metrics related to SDN and NFV layers and components. We will analyze deeper the vulnerabilities and challenges of NFV to understand how to adapt our Self-Healing framework. The goal is to tackle those vulnerabilities on SDN and NFV through a proactive Self-Healing framework.

## VI. REFERENCES

[1] FP7 Programme Project Mobile and wireless communications Enablers for the Twenty-twenty Information Society (METIS)

[2] J. O. Kephart and D. M. Chess., "The Vision of Autonomic Computing" in IEEE Computer, Vol. 36, No. 1, pp. 41-50, 2003.

[3] ONF, "SDN Architecture Overview v1.0, Dec 2013.

[4] Updated White paper on "Network Functions Virtualisation".

[5] A. Basta et al., "A Virtual SDN-Enabled LTE EPC Architecture: A Case Study for S-/P-Gateways Functions," Proc. 2013 IEEE SDN for Future Networks and Services, Trento, Italy, Nov. 2013.

[6] Xin Jin et al., "CellSDN: Software-Defined Cellular Core Networks".

[7] J. Batalle et al., "On the Implementation of NFV over an OpenFlow Infrastructure: Routing Function Virtualization", Proc. 2013 IEEE SDN for Future Networks and Services, Trento, Italy, Nov. 2013.

[8] K. Hyojoon et al., "CORONET: Fault tolerance for Software Defined Networks", Network Protocols (ICNP), 2012 20th IEEE International.

[9] P. Fonseca, et al., "A replication component for resilient OpenFlow-based networking", NOMS 2012.

[10] P. Sharma et al., "Enhancing network management frameworks with SDN-like control", IM 2013, 27-31 May 2013.

[11] S. Sharma et al., "Fast failure recovery for in-band OpenFlow networks", DRCN 2013.

[12] ETSI, "An Architectural Reference Model for Autonomic Networking, Cognitive Networking and Self-Management".

[13] 3GPP SON, Available at: http://www.3gpp.org/technologies/keywords-acronyms/105-son.

[14] 5G-PPP, "The 5G infrastructure Public Private Partnership", Available at: http://5g-ppp.eu.

[15] J. Sanchez, I. Grida Ben Yahia, N. Crespi, "Self-healing Mechanisms for Software Defined Networks". AIMS 2014.